\documentclass[twocolumn,showpacs,amsmath,amstex,amssymb,mathfonts,nobalancelastpage*]{revtex4}
\usepackage{graphicx,bm}
\usepackage{epstopdf}
\usepackage{color}

\begin{document}
\title{Singular Behavior of Anderson Localized Wavefunctions for a Two-Site Model}
\author{S. Johri$^{1}$ and R. N. Bhatt$^{1,2}$}
\affiliation{$^1$ Department of Electrical Engineering, Princeton University, Princeton, NJ 08544}
\affiliation{$^2$ Princeton Center for Theoretical Science, Princeton University, Princeton, NJ 08544}\textbf{}
\begin{abstract}
We show analytically that the apparent non-analyticity discovered recently in the inverse participation ratio (IPR) of the eigenstates in Anderson's model of localization is also present in a simple two-site model, along with a concurrent non-analyticity in the density of states (DOS) at the same energy. We demonstrate its evolution from two sites to the thermodynamic limit by numerical methods. For the two site model, non-analyticity in higher derivatives of the DOS and IPR is also proven to exist for all bounded distributions of disorder. 
\end{abstract}
\pacs{71.23.An, 71.30.+h, 72.80.Ng}
\date{\today}
\maketitle

\section{Introduction}
In a recent paper\cite{johri-bhatt}, we reported numerical studies of the Anderson model of localization\cite{anderson} in the insulating regime. We studied, in particular, the ensemble-averaged inverse participation ratio (IPR), $I(E)$, which characterizes the extent of eigenstates at a given energy $E$. We found that $I(E)$ exhibits a sharp, apparently singular, behavior as a function of $E$, at a specific energy $E$, which lies in the insulating phase, but is clearly distinct from the true band edge. This sudden change in behavior was interpreted in terms of a sharp transition from a regime comprising for the most part of typical, Anderson-localized states, to one comprising almost exclusively states involving resonance between two (or more) sites. Such states involving resonance among a large number of sites are known in the literature as Lifshitz states\cite{kramer}, \cite{lifshitz}. 

This aforementioned behavior was seen in one, two and three dimensions for bounded distributions of on-site energies (e.g. the uniform distribution of width $w$ with a constant probability density between the limits $\pm w/2$.),  but not for unbounded distributions ({\it e.g.} Gaussian). The apparent singular behavior could be observed most clearly for disorder exceeding a certain, moderate value (disorder width $w = 3.8$ for the uniform distribution in $d = 1$), and appeared to persist for extremely large disorder, where most states at the energies of interest are localized on a few sites. This provides motivation for studying various quantities analytically for the Anderson model defined on a small, finite number of sites. The simplest of these exhibiting the difference between typically localized and resonant states is the two-site model, which, as we show below, can be solved analytically.

The solution for the two-site model shows that for all bounded distributions defined in the interval $(-w/2, w/2)$, both the ensemble averaged IPR, $I(E)$, and the density of states $\rho(E)$, exhibit singular behavior at the energies $E=\pm E_2$ whose value is related to the disorder width $w$ by the simple algebraic expression $E_2 = \sqrt {w^2/4 + 1}$. The exact nature of the singularity depends on the form of the distribution characterizing the diagonal disorder of the Anderson model. In particular for the uniform distribution of Anderson's original paper\cite{anderson}, with a step discontinuity at $-w/2$ and $+w/2$, both quantities show a slope discontinuity. This is rather similar to what is seen numerically for the thermodynamic system\cite{johri-bhatt}, though the energy at which the apparent singular behavior is seen for the thermodynamic system differs from that of the two-site model. Nevertheless, the reason for the discontinuity in slope is found to be the same in the two-site model, namely the loss of certain kind of states beyond a critical energy\cite{johri-bhatt}, so it is not unreasonable to conjecture that the two phenomena are related.

For other bounded distributions, with power-law thresholds, the singularity for the two-site model appears in a higher order derivative, while for integrable, inverse power law thresholds (where the distribution diverges at the edges), the {\it derivatives} of $\rho(E)$ and $I(E)$ diverge at $E=\pm E_2$.

The plan of the paper is as follows. In Section II, we define the n-site Anderson model on a ring ({\it i.e.}, with periodic boundary conditions). Since all states of the Anderson model are known to be localized in $d = 1$, this set of models are representative of the localized phase, and also of the true thermodynamic model for sufficiently large disorder. We then specialize to the case of two sites and present complete results for the ensemble averaged density of states and the inverse participation ratio for the case of Anderson's uniform distribution. The details of the calculation, provided in Appendix A, make it clear that a bounded distribution is necessary to obtain singular points for the two-site model. We then consider other bounded distributions with power law singularities, and demonstrate the singular behavior of appropriate derivatives of $\rho(E)$ and $I(E)$. Details of the derivation are provided in Appendix B. In Section III, we present our results for $\rho(E)$ and $I(E)$ for the n-site Anderson model with a uniform distribution obtained numerically for two values of $w$, representing moderate and large disorder, and show how these quantities evolve with n. Finally, in Section IV, we summarize our results.


\section{Anderson Model and Symmetric, Bounded Probability Distributions}

\begin{figure}
	\centering
	\includegraphics[width=\columnwidth]{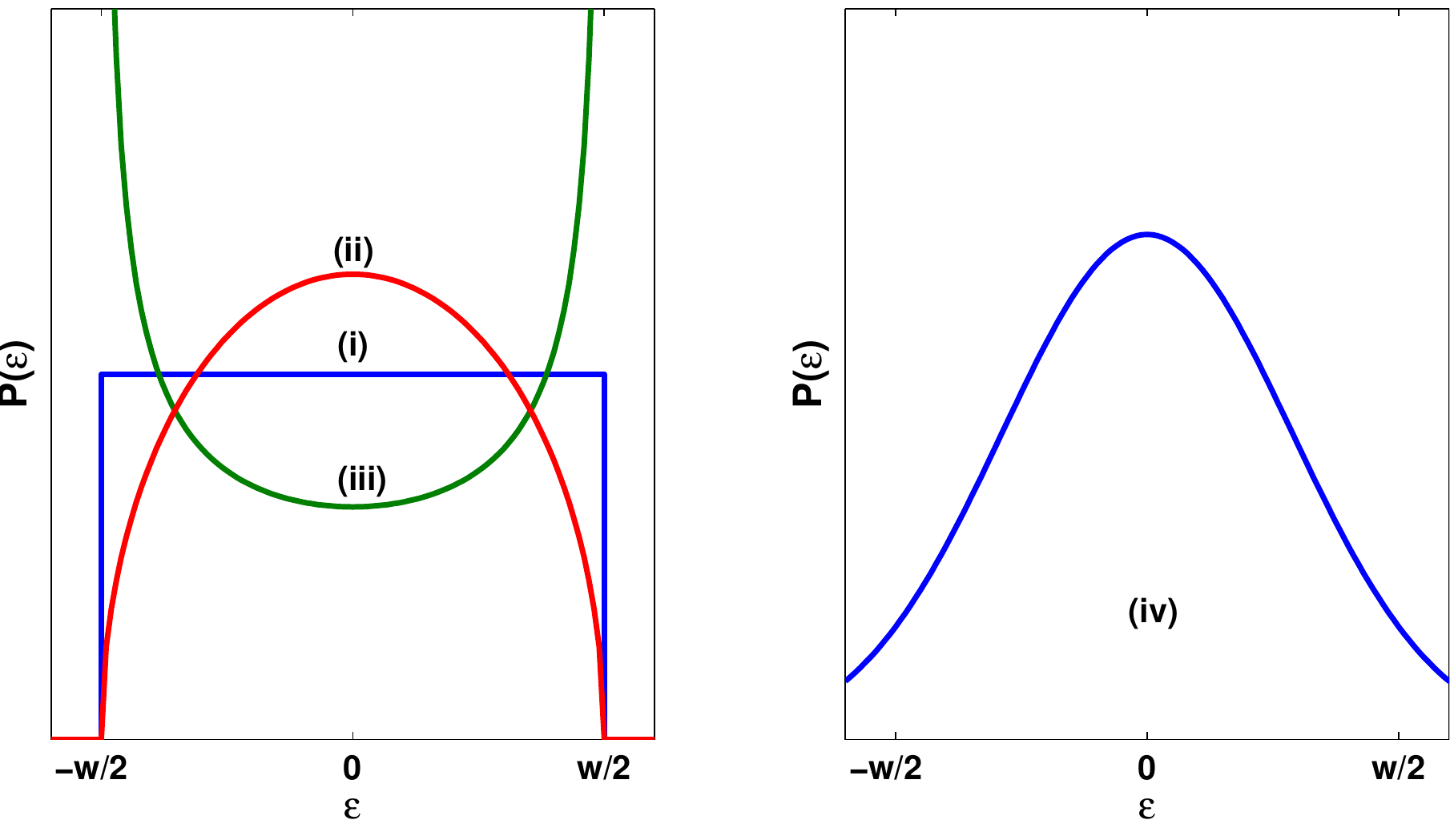}
	\caption{Some normalized distributions of the on-site energy considered in this paper. On the left are three bounded distributions: (i) Uniform, $P(\epsilon)=1/w$, (ii) \& (iii) have power law edges, $P(\epsilon)=\frac{2 \Gamma(3/2+\lambda)}{\sqrt{\pi}w\Gamma(1+\lambda)}(1-4\epsilon^2/w^2)^\lambda$ for $\lambda=1/2$ and $\lambda=-1/2$ respectively. On the right is the unbounded Gaussian distribution.}
	\label{fig:dist}
\end{figure}

The Anderson model Hamiltonian is given by:
\begin{equation}\label{eq:ham}
H = \sum_{i}\epsilon_i |i><i| + V \sum_{<i,j>} |i><j|						
\end{equation}
where $|i>$ are states localized on sites $i$ of a simple hypercubic lattice in $d$-dimensions, and $i$, $j$ are nearest neighbours. The onsite energies $\epsilon_i$ are independent random variables taken from a specified distribution $P(\epsilon)$. By definition, $\int_{-\infty}^{\infty}P(\epsilon) d\epsilon=1$. We take $P(\epsilon)$ to be non-zero only between $\pm w/2$. Within these limits, the distribution can have any form as long as it is continuous and symmetric in energy. Examples are the uniform, semicircular and inverse-power law distributions (Fig.\ref{fig:dist} (i) - (iii)). Later in this section, we will briefly discuss what happens when $P(\epsilon)$ is not bounded (e.g. Gaussian distribution (Fig.\ref{fig:dist} (iv))). In the thermodynamic limit, for symmetric $P(\epsilon)$ and $d$-dimensional hypercubic lattices which are bipartite (including the nearest neighbour chain that we consider here), properties are symmetric around $E=0$. The symmetry in energy also holds for finite size systems that have an even number of sites. 

The hopping $V$ is assumed to have the same non-zero value for all nearest neighbours, and zero for non-nearest neighbours. In this paper, we use $V=-1$ to define our unit of energy. In one-dimension, which is what we consider in this paper, all eigenstates in the Anderson model are known to be localized\cite{mott} for non-zero $w$. The localization length decreases as the disorder, parametrized by the range of possible values of $\epsilon_i$ (e.g. the width of its probability distribution $w$), increases. 

The IPR for a wavefunction $\Psi=\sum_{i}a_i |i>$ is defined as 
\begin{equation}\label{eq:defipr}
I_{\Psi}=\frac{\sum_{i} {|a_i|^4}}{\left(\sum_{i}{|a_i|^2}\right)^2}						
\end{equation}
It is inversely proportional to the width (support) of the wavefunction. For localized eigenstates, the IPR goes to a constant as the system size tends to infinity. As detailed in our previous paper, in the Anderson model which uses a uniform distribution, the IPR has a local (non-zero) minimum at the centre of the band, rises to a maximum and turns around sharply in an apparently non-analytic manner and goes to zero at the band edges $E=\pm(w/2+2)$.

We now consider the Anderson model for finite lattices. For two sites (1,2), the Hamiltonian of Eq. \ref{eq:ham} reduces to
\begin{equation}\label{eq:ham_two}
H = \epsilon_1 |1><1|+\epsilon_2 |2><2| - |1><2| -	 |2><1|						
\end{equation}
i.e. a $2\times2$ matrix, which one can easily solve for arbitrary $\epsilon_1$ and $\epsilon_2$. The eigenvalues are
\begin{equation}
E_{\pm}=\frac{x}{\sqrt{2}} \pm \sqrt{\frac{y^2}{2}+1}
\end{equation}
where $x=\frac{\epsilon_{1} + \epsilon_{2}}{\sqrt{2}}$ and $y=\frac{\epsilon_{1} - \epsilon_{2}}{\sqrt{2}}$. The corresponding eigenstates can be written as $\alpha|1>+\beta|2>$ where
\begin{equation}
\bigg(\frac{\beta}{\alpha}\bigg) _{\pm}=\frac{y}{\sqrt{2}}\mp \sqrt{\left(\frac{y}{\sqrt{2}}\right)^2+1}
\end{equation}
The IPR for both wavefunctions is the same:
\begin{equation}\label{eq:ipr}
I_{\pm}=1-\frac{1}{1+y^2/2}				
\end{equation}
which is further independent of x.

The random variables $\epsilon_1$ and $\epsilon_2$ have a probability distribution $P(\epsilon)$. The ensemble averaged DOS ($\rho$) and IPR ($I$) are given by:
\begin{eqnarray}\label{eq:dos}
\rho(E) &=& \frac{1}{2}\int_{-\infty }^{\infty} \int_{-\infty}^{\infty} d \epsilon_1  d \epsilon_2  P(\epsilon_1) P(\epsilon_2) \times {}
\nonumber\\
&& {} [\delta(E-E_1)+\delta(E-E_2)]
\end{eqnarray}
\begin{eqnarray}\label{eq:ipr}
I(E) & = & \frac{1}{\rho(E)}\frac{1}{2}\int_{-\infty}^{\infty} \int_{-\infty}^{\infty} d \epsilon_1  d \epsilon_2 P(\epsilon_1) P(\epsilon_2)\times{}
\nonumber\\
&& {}  [\delta(E-E_1)+\delta(E-E_2)]\times{}
\nonumber\\
&&{}\bigg(1-\frac{1}{2(1+\frac{y^2}{2})}\bigg)
\end{eqnarray}
Transforming the variables to x and y makes the integrals easier to evaluate. The Jacobian is unity, giving $d\epsilon_1 d \epsilon_2 = dxdy$.  

For $w=0$, the DOS will be two delta functions at $E=\pm 1$. For $0<w<2$, the DOS goes to zero for $w/2-1<E<-w/2+1$ (see Appendix A). Since the two-site model is expected to be a good approximation for the infinite model only at large $w$, we take $w>2$ from now on.

Under these conditions, the integrands in Eq. \ref{eq:dos} and Eq. \ref{eq:ipr} are non-zero only over the square shown in Fig.1. As shown in Appendix A, the arguments of the delta functions in the integrals are just two branches of a hyperbola symmetric about the x-axis, $(x/\sqrt{2}-E)^2-y^2/2=1$. At $E=0$, it is also symmetric about the y-axis. As $E$ increases (decreases), the hyperbola moves right (left) as shown in Fig \ref{fig:sqhyper} of Appendix A. Any ensemble averaged property of the wavefunctions can be calculated as a function of energy by doing the relevant line integral along the length of the hyperbola lying inside the square.

\begin{figure}
	\centering
	\vspace{-50 pt}
		\hspace*{-70pt}
	\includegraphics[width=1.5\columnwidth]{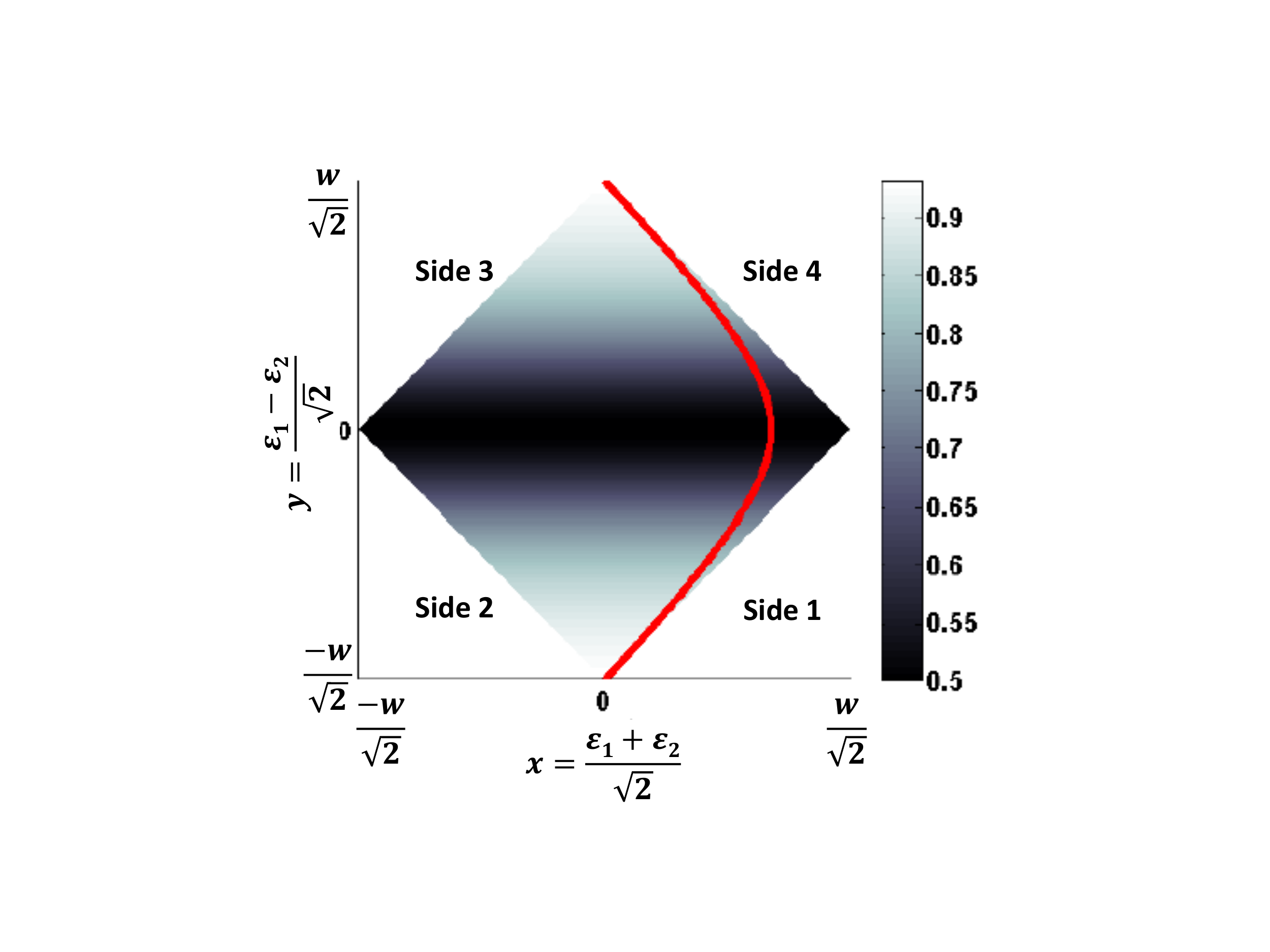}
		\vspace{-50pt} 
	\caption{Allowed values of x and y lie inside the square. Colorbar shows variation of IPR over square. The red curve shows the left branch of the hyperbola when it is tangent to sides 1 and 4 of the square.}
	\label{fig:rot_sq}
\end{figure}

The simplest bounded probability distribution is one in which the $P(\epsilon)$ is a constant between $-w/2$ and $w/2$. This is also the distribution considered in Anderson's original paper\cite{anderson}. The complete expressions for the DOS and IPR when the probability distribution is uniform are given in Table \ref{table:dos_ipr_expr} and the derivation is provided in Appendix A. In Fig \ref{fig:w2pt25} and Fig \ref{fig:w3}, we plot (as solid lines) the analytical DOS and IPR for 2 sites as a function of energy for $w=4.5$ and $w=6$ respectively which agree with our numerical results shown as rightward-pointing triangles. Going from the centre to the edge of the band, the slope of the DOS and IPR discontinuously changes sign at $E=\pm E_2$. This can be understood as a loss of typical ``Anderson'' type states as follows.


\begin{table}{
\hspace{-10pt}
\renewcommand{\arraystretch}{2.5}
    \begin{tabular}{ | c | c | c | p{5 cm} |}

   \hline
     Energy Range &  $\rho(E)$ & $I(E)$ \\ \hline
    $-\frac{w}{2}-1\rightarrow -\sqrt{\frac{w^2}{4}+1}$ & $2\alpha$ & $1-\frac{1}{\alpha}\tan^{-1}(\frac{\alpha}{2})$  \\ \hline
    $-\sqrt{\frac{w^2}{4}+1}\rightarrow -\frac{w}{2}+1$ & $2\beta$ & $1-\frac{1}{\beta}\tan^{-1}(\frac{\beta}{2})$ \\ \hline
    $-\frac{w}{2}+1\rightarrow \frac{w}{2}-1$ & $2(\alpha+\beta)$ & $1-\frac{\tan^{-1}(\frac{\beta}{2})+\tan^{-1}(\frac{\alpha}{2})}{\alpha+\beta}$ \\ \hline
    $\frac{w}{2}-1\rightarrow \sqrt{\frac{w^2}{4}+1} $ & $2\alpha$ & $1-\frac{1}{\alpha}\tan^{-1}(\frac{\alpha}{2})$ \\ \hline
    $\sqrt{\frac{w^2}{4}+1}\rightarrow \frac{w}{2}+1$ & $2\beta$ &$1-\frac{1}{\beta}\tan^{-1}(\frac{\beta}{2})$\\ \hline
        \end{tabular}
        
    \caption{Analytical expressions for DOS and IPR over the whole band for the uniform distribution when $w>2$. Here, $a=E+w/2$, $b=E-w/2$ and $\alpha= \frac{a^2-1}{a}$, $\beta=\frac{1-b^2}{b}$.}
    \label{table:dos_ipr_expr}}
\end{table}

The IPR is an average over all the states lying on the hyperbola. Each point in the square will contribute to the average at two different energies, once when it lies on the left branch of the hyperbola and once when it lies on the right. The states with the highest IPR lie on the upper and lower vertices of the square. These states occur when $|y|$ is maximum, i.e. $\epsilon_1$ and $\epsilon_2$ are far from each other. Near the center of the square, $\epsilon_1$ is close to $\epsilon_2$ --- wavefunctions arising from these configurations can be called `resonant' states.  

At $E=E_2$, the left branch of the hyperbola becomes tangent to the sides 1 and 4 of the square as shown in Fig. 1. In this position, the maximum length of the hyperbola passes through the high IPR region. The average IPR will thus be maximum in this region. However, at any energy close to this, the hyperbola shifts and the high-IPR states start being lost leading to a decrease in the average IPR. The rate of decrease is different when the shift is left or right, leading to a discontinuity in the slope $\frac{dI(E)}{dE}$ of the average IPR. When the hyperbola moves left ($E$ decreases), it still passes through a large number of high IPR states. When it moves right ($E$ increases), it loses the high-IPR states much faster, and passes through states for which $\epsilon_1+\epsilon_2$ is large and $\epsilon_1-\epsilon_2$ is small, implying that both sites have energies close to $w/2$. These type of resonant states which occur near the edge of the band and arise from clusters of sites which have energies close to one of the band edges, are called Lifshitz states. They are separated here from the rest of the band by the non-analytic point in the IPR.

The softer discontinuity in the DOS is only due to a decrease in the total number of states lying on the hyperbola, irrespective of whether they are regular Anderson localized or resonant states.



For a non-constant (but still bounded) probability distribution, the discontinuity may be present in higher derivatives of the DOS and IPR. As shown in Appendix B, the expression for $(d^n\rho/dE^n)$ or $(d^n I/dE^n)$ will have terms of the form $(d^{m_1}P(\epsilon_1)/d\epsilon_1^{m_1}) (d^{m_2}P(\epsilon_2)/d\epsilon_2^{m_2})$ with $(m_1+m_2)$ ranging from $0$ to $(n-1)$. If any of these terms are non-zero at the edge, there will be a discontinuity in that derivative at $E=\pm E_2$. If any of them diverge at the edge, then the corresponding $(d^n\rho/dE^n)$ and $(d^n I/dE^n)$ will also diverge at $E=\pm E_2$. Thus, if a probability distribution $P(\epsilon)\rightarrow (1-4\epsilon^2/w^2)^\lambda$ as $\epsilon\rightarrow w/2$ with $-1<\lambda<0$, then all the derivatives of DOS and IPR will be infinite at $E=\pm E_2$. For $\lambda$ positive, the divergence will be in a higher derivative.

From the above discussion, it is easy to see that a non-bounded probability distribution like the Gaussian, for which the boundaries of the square lie at infinity, will not exhibit any discontinuities in the 2-site model.

\section{Numerical Results}

\begin{figure}
	\centering
	\includegraphics[width=\columnwidth,height=2\columnwidth]{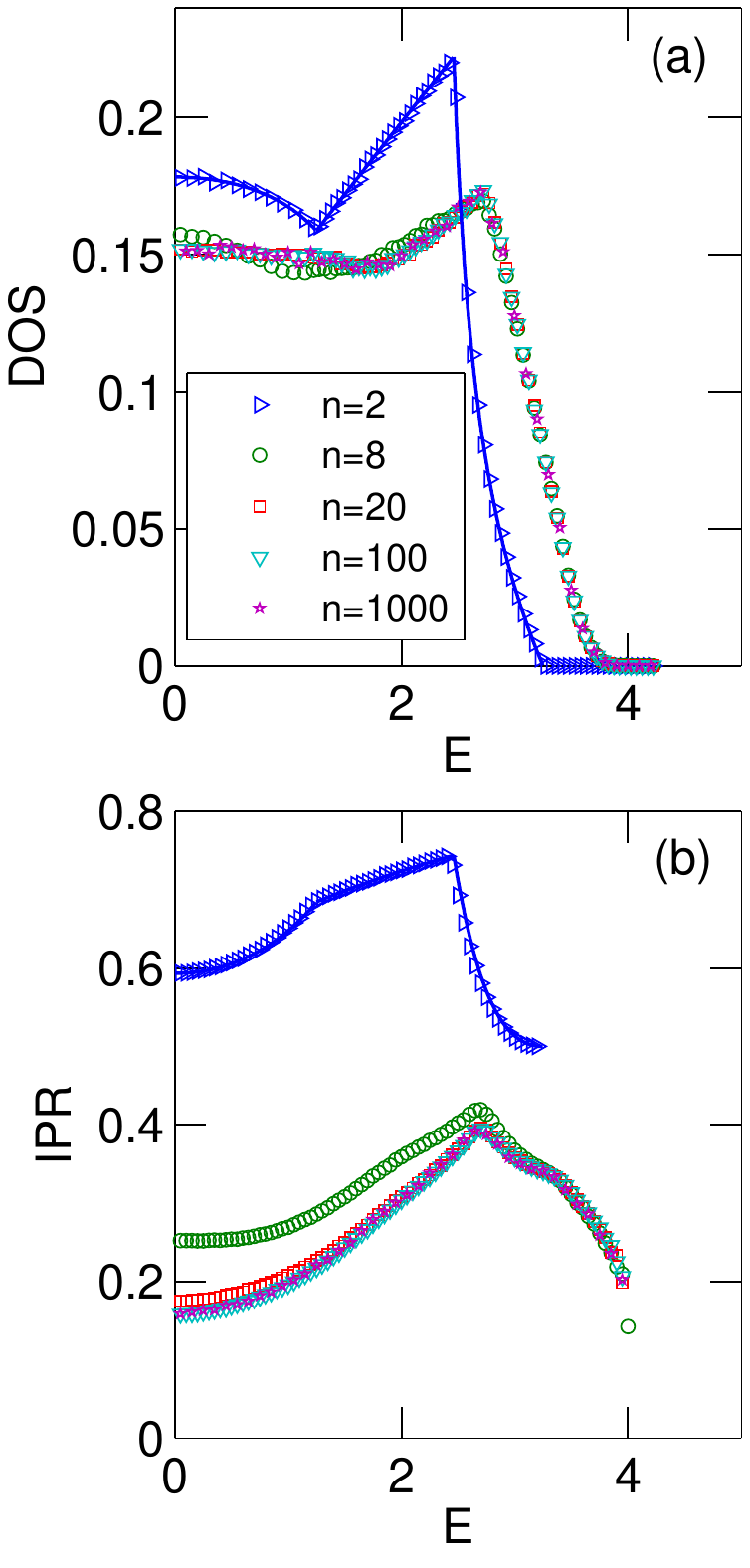}
	\caption{Evolution of DOS and IPR with increasing system size for $w=4.5$. Solid line shows analytical result for n=2.}
	\label{fig:w2pt25}
\end{figure}

\begin{figure}
	\centering
	\includegraphics[width=\columnwidth,height=2\columnwidth]{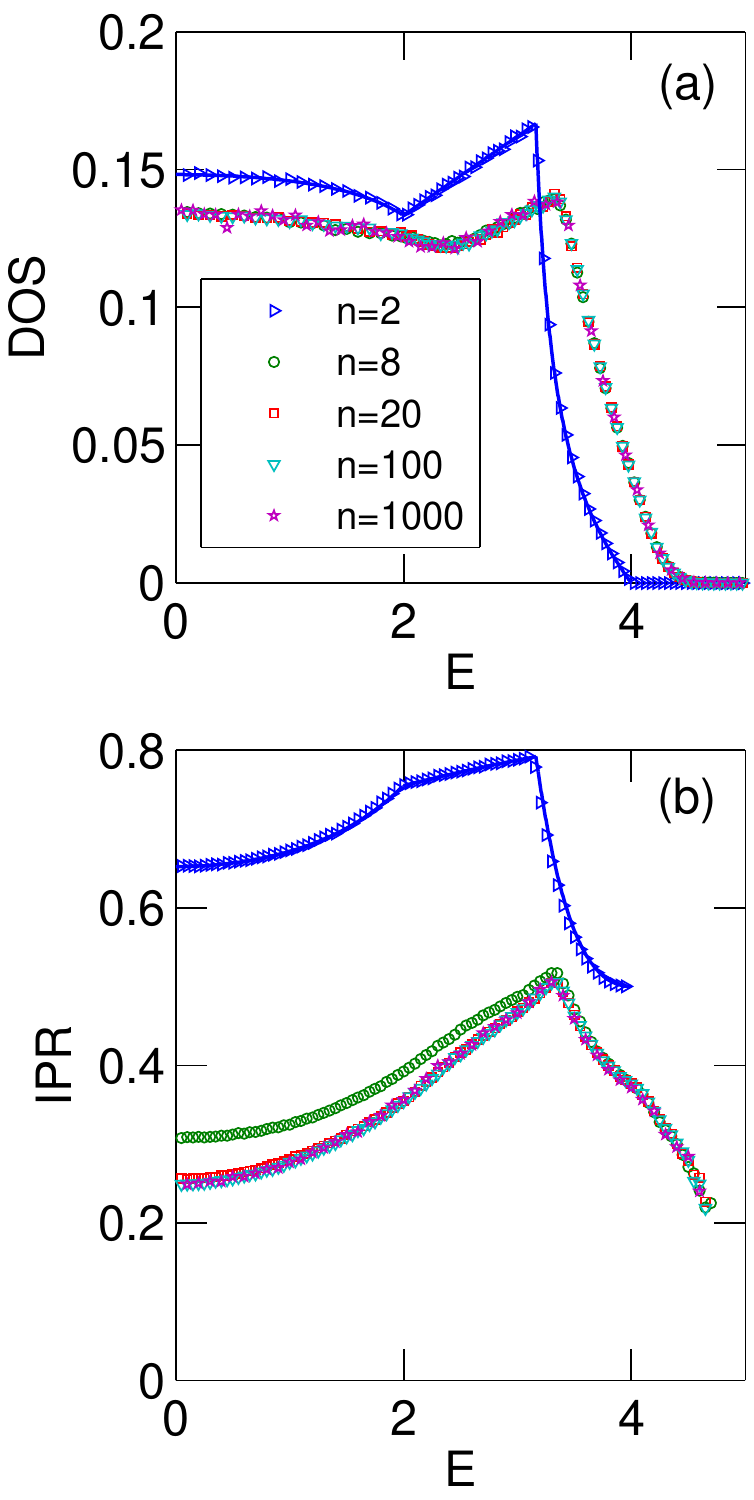}
	\caption{Evolution of DOS and IPR with increasing system size for $w=6$. Solid line shows analytical result for n=2.}
	\label{fig:w3}
\end{figure}

In Fig \ref{fig:w2pt25} and Fig \ref{fig:w3}, we plot with varying symbols numerical results for $n$-site chains for typical values of disorder, $w=4.5$ and $6$. These typical cases show how the DOS and IPR converge to their thermodynamic limits relatively quickly as $n$ is increased, as expected since we are in the localized regime. Larger values of $w$ (larger disorder) require smaller size systems to reach the thermodynamic limit within a given precision as the localization length decreases with increasing disorder. Thus, for $w=4.5$, a lattice size of 100 is required to obtain results significantly closer to the thermodynamic limit than the size of the symbols, whereas for $w=6$, only 20 sites are required. The discontinuity in the derivative of the IPR survives and seems to grow sharper in the thermodynamic limit, whereas that for the DOS seems to flatten out a bit. There also seems to be an additional feature in the IPR beyond the maximum. This may be the start of a transition into another regime of resonant states, e.g. from 2 sites to 3 sites. However, this is hard to confirm numerically. Interestingly, as would be expected if the above conjecture were true, it is not present in the two-site model.

\begin{figure}
	\centering
		\hspace*{-1000pt}
	
	\includegraphics[width=\columnwidth]{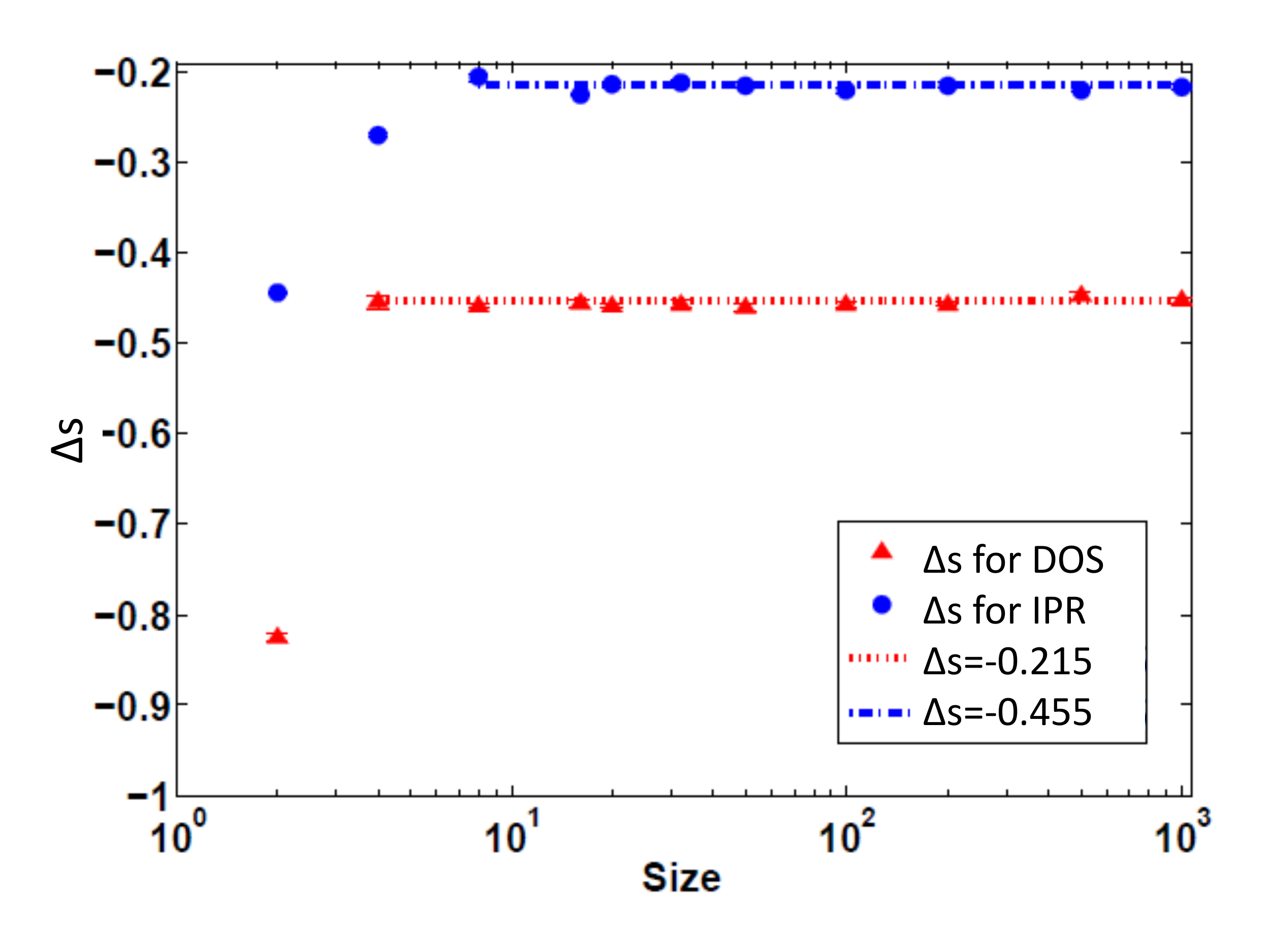}
		\vspace{-15pt} 
	\caption{$\Delta s = (s_{left} - s_{right})$ as a function of system size for $w=6$ for DOS (triangles) and IPR (circles).}
	\label{fig:diffslope}
\end{figure}

To measure the evolution of the non-analytic behavior with size, the curves on either side of the roughly estimated maxima of the DOS and IPR were fit with quadratic functions up to the points where additional kinks occur in the curve. The position of the maxima, $E_{\rho,1}$ and $E_{I,1}$ were then determined from the intersection of these fits. The difference of the slopes $s_{left}$ and $s_{right}$ of the fits near the maxima were used to quantify the discontinuity. ($\Delta s=s_{left}-s_{right}$) is shown to reach a constant value with increasing size in Fig \ref{fig:diffslope}, suggesting that the singularity survives in the thermodynamic limit.

\begin{figure}
	\centering
	\includegraphics[width=\columnwidth]{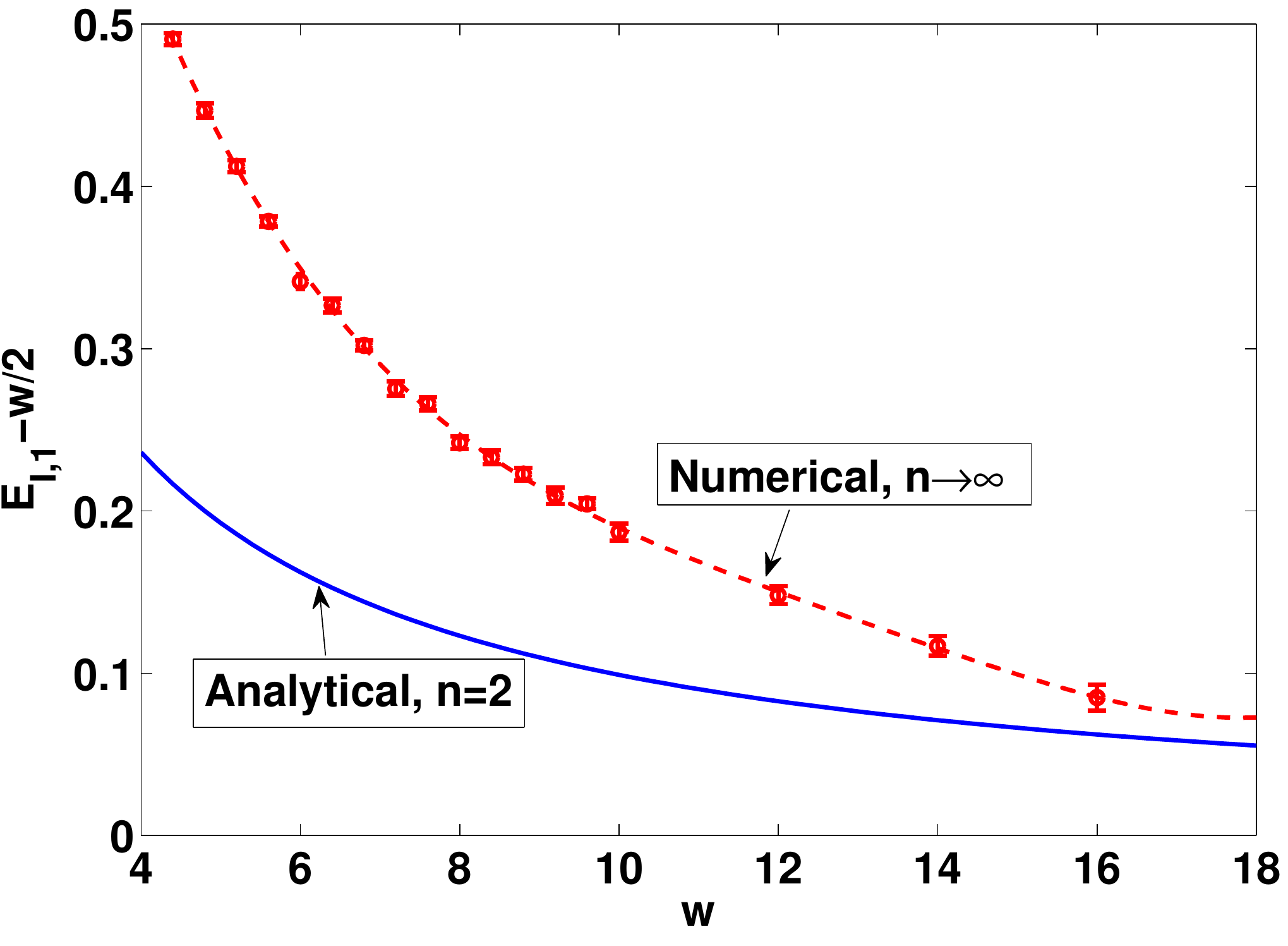}
	\caption{Comparison of value of energy at which singularity occurs for 2 sites (blue solid line) and the large size limit (red circles). The red dotted line is a fit to the circles. The leading term $w/2$ is subtracted from the energy to amplify the difference.}
	\label{fig:maxpos}
\end{figure}

Fig \ref{fig:maxpos} compares the analytically determined point of discontinuity ($E=\sqrt{w^2/4+1}$)in the 2-site case to the numerically determined one in the infinite size limit. For large disorder, the 2-site problem seems to be a good approximation to the infinite size limit, whereas for smaller $w$ the deviation is larger. This is to be expected since increasing disorder leads to smaller localization lengths.

\section{Conclusions}
In conclusion, we have shown by direct calculation how the non-analyticity in the DOS and IPR arises for the 2-site model for bounded disorder. Further, we have shown the evolution with number of sites for the case of uniform disorder and found an apparent singularity, at least for IPR, surviving in the thermodynamic limit. Our numerical studies on different lattice sizes suggest that there is an exact energy at which high IPR states are lost and that the position as well as the magnitude of the singularity approach a limit as $n \rightarrow \infty$. It is at least clear that the singularity arises due to the bounded nature of the disorder distribution, which leads to loss of high-IPR wavefunctions at a particular energy in the band like in the 2-site model. The non-analyticity appears to be present for all lattice sizes and we surmise that it survives in the thermodynamic limit. The question of whether it is possible to show that the non-analyticity persists in the thermodynamic limit by analytical means remains unanswered at present.

We also remark that studying a 3-site model may allow us to determine if the shoulders seen in the IPR curves in Fig. \ref{fig:w2pt25} and Fig. \ref{fig:w3} are due to transition from 2 to 3 resonant sites.

After this work was completed, we received news of parallel work by Ujfalusi \& Varga \cite{varga} which addresses some of the issues discussed in this paper.

This work was supported by DOE grant DE-SC20002140.

\section{Appendix A}
The integrands in Eq. \ref{eq:dos} and Eq. \ref{eq:ipr} contain delta functions in a 2 dimensional space. Using the identity for $n$- dimensions:
\begin{equation}
\int{\delta(g(\vec{r})) f(\vec{r}) d \vec{r}^n}= \int_{g(\vec{r})=0}{\frac{d \sigma(\vec{r})}{|\vec{\nabla}g|} f(\vec{x})} 	
\end{equation}
where the integral on the right is over the hypersurface $g(\vec{r})=0$, in one dimension lower than the one on the left. In our case $g(\vec{r})=E-\frac{1}{\sqrt{2}}(x \pm \sqrt{y^2 + 2})$ i.e. the equation of a hyperbola for fixed $g(\vec{r})$, here $g(\vec{r})=0$. When $E=0$, the hyperbola is symmetric about the $y$-axis, and when $E$ is non-zero, the hyperbola is displaced along the $x$-axis. It is always symmetric about the $x$-axis.

Because of the delta function and the bounds on $x$ and $y$, we effectively have to calculate a line integral along the portion of the hyperbola which lies inside the square. The hyperbola can be parametrized by $x=\sqrt{2} (\sec (t) + E)$, $y=\sqrt{2} \tan (t)$. Thus, we have $|\vec{\nabla}g|=\sqrt{\frac{y^2+1}{y^2+2}}=\sqrt{\frac{2\tan^2(t)+1}{2 \sec^2(t)}}$ and $d\sigma=\sqrt{2 \sec^2(t) \tan^2(t)+2 \sec^4 (t)}dt = \sqrt{2} \sec(t) \sqrt{\tan^2(t)+ \sec^2(t)}$. 
The intersection points of the sides of the square (numbered as in Fig.1) with the hyperbola are: $y_3=\frac{(\frac{w}{2}+E)^2-1}{\sqrt{2}(\frac{w}{2}+E)}$, $y_4=\frac{(\frac{w}{2}-E)^2-1}{\sqrt{2}(\frac{w}{2}-E)}$, $y_1=-y_4$ and $y_2=-y_3$. 

\begin{figure}
	\centering
	\includegraphics[width=4.8cm,height=10 cm]{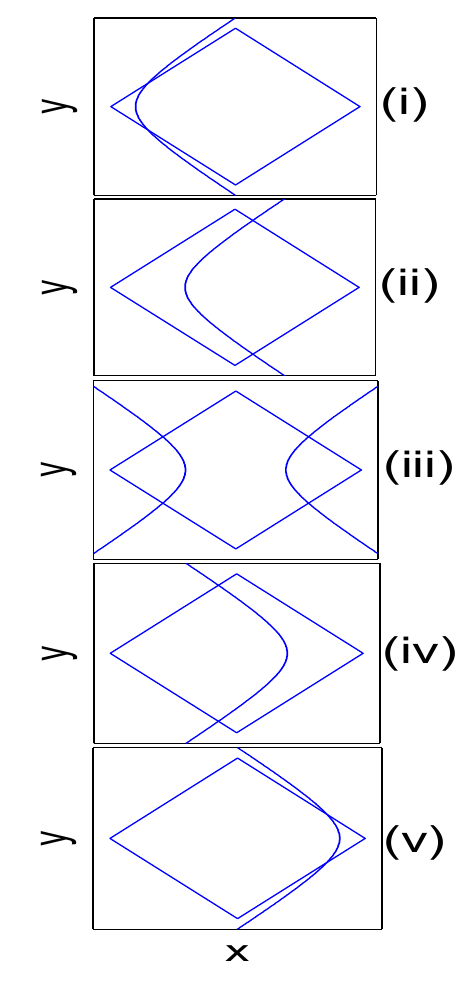}
	\caption{Hyperbola $E=\frac{1}{\sqrt{2}}(x \pm \sqrt{y^2 + 2})$ moves to the right as $E$ increases. Figures (i)-(v) depict sequentially the five cases in Table \ref{table:dos_ipr_expr}. Allowed values of $x$ and $y$ lie inside the square.}
	\label{fig:sqhyper}
\end{figure}

The five possible situations for $w>2$ are illustrated in Fig. \ref{fig:sqhyper}. The size of the square depends only on $w$. The position of the hyperbola depends only on the energy. For $w<2$, the density of states goes to zero in the middle case (Fig. \ref{fig:sqhyper}(iii)) since the square is small enough to lie between the two branches of the hyperbola without intersecting either of them.

In cases (i) and (ii), only the right branch of the hyperbola ($E_-$) falls inside the square, for (iv) and (v), we need to account for only the left branch ($E_+$), and in case (iii), both branches are required. For positive energies, the point which gives rise to the change in sign of the slopes of average DOS and IPR is at the transition from case (iv) to case (v) (There will also be a smaller discontinuous change in the slopes when transitioning from (iii) to (iv)).

The integrand is independent of $E$ and the dependence on $E$ is provided by the limits of the integration. The average IPR and DOS for the uniform distribution are
\begin{equation}
	\rho(E)= \int_{-y_0(E)}^{y_0(E)}{2 \sec^2(t) dt} 
\end{equation}
Similarly,
\begin{equation}
I(E)= \frac{1}{\rho(E)}\int_{-y_0(E)}^{y_0(E)}{2 \sec^2(t) \bigg(1-\frac{1}{2 \sec^2(t)}\bigg)}dt
\end{equation}
with limits of the above integrals given by the intersection points and using one or both branches of the hyperbola as appropriate. The final expressions are given in Table \ref{table:dos_ipr_expr}.

The transitions between case (i) to case (ii) and case (iv) to case (v) occur at $E=\pm E_2$. At these points, $y_3=y_4 =\frac{w}{\sqrt{2}}$. From Eq. 2, $x=0$ at his point. This corresponds to the situation $\epsilon_1=w/2$ and $\epsilon_2=-w/2$ or vice versa, that is, the two sites lie at opposite ends of the probability distribution.

\section{Appendix B}
The DOS (and IPR) contains contributions $\rho_1$ and $\rho_2$ from the 2 branches of the hyperbola. In this section, the analysis is carried out only for $\rho_1$. The other term can be calculated in an analogous manner.

For any symmetric, bounded and continuous probability distribution $P(\epsilon)$, the DOS due to the left branch of the hyperbola can be expressed in x and y coordinates as:
\begin{eqnarray}\label{eq:dos_gen}
	\rho_1(E) & = & \frac{1}{2}\int dx dy P\left(\frac{x+y}{\sqrt{2}}\right)P\left(\frac{x-y}{\sqrt{2}}\right)\times {}
	\nonumber\\
	&& {}  \delta\left(E-\frac{x}{\sqrt{2}} - \sqrt{\frac{y^2}{2}+1}\right)
	\nonumber\\
\nonumber &=& \frac{1}{2}\int_{-y_0(E)}^{y_0(E)} \frac{dy}{1/\sqrt{2}} P\left(\frac{x(E,y)+y}{\sqrt{2}}\right)\times\\
& &P\left(\frac{x(E,y)-y}{\sqrt{2}}\right)
\end{eqnarray}
In the integrand $x$ is now a function of $E$ and $y$: $x=\sqrt{2}E+\sqrt{y^2+2}$. $P\left(\frac{x+y}{\sqrt{2}}\right)=P(\epsilon_1)$ and $P\left(\frac{x-y}{\sqrt{2}}\right)=P(\epsilon_2)$. The limits $y_0(E)$ depend on the width of the probability distribution. 

Differentiating Eq.\ref{eq:dos_gen} with respect to $E$,
\begin{align}
\frac{d \rho_1(E)}{dE}&= \sqrt{2}\left\{ \left[\int_0^{y_0} {dy \frac{\partial (P(\epsilon_1) P(\epsilon_2))}{\partial E}}\right] \right.\nonumber\\
&\qquad \left. {}+ [P(\epsilon_1) P(\epsilon_2)]_{y=y_0} \frac{d y_0}{dE}\right\}
\end{align}
Consider the left and right limits of $\frac{d \rho_1(E)}{dE}$ at $E=-\sqrt{\frac{w^2}{4}+1}$  when the probability distribution is bounded between $-w/2$ and $w/2$. $y_0=y_3$ (intersection of the hyperbola with sides 2 and 3 of the square) for the left limit and $y_0=y_4$ (intersection with sides 1 and 4) for the right limit. The first term (the integral) remains the same for both since at the critical energy $E=-\sqrt{\frac{w^2}{4}+1}$, $y_3=y_4=w/\sqrt{2}$. Now consider the second term. $\frac{d y_3}{dE} \neq \frac{d y_4}{dE}$, $P(\epsilon_1)=P(w/2)$ and $P(\epsilon_2)=P(-w/2)$, i.e. equal to their value at the edges. For the uniform distribution, these are non-zero, and therefore the first derivative of $DOS$ is discontinuous. 

Similarly, we can calculate the second derivative:

\begin{eqnarray}
\frac{d^2 \rho_1(E)}{dE^2} &=& \sqrt{2} \left(\int_0^{y_0} {dy \frac{\partial^2 (P(\epsilon_1) P(\epsilon_2))}{\partial E^2}}\right) {}
\nonumber\\
&& {} + \left(\frac{d (P(\epsilon_1) P(\epsilon_2))_{y=y_0}}{d E}\right) \frac{d y_0}{dE}
\nonumber\\
&& {} + \left(\frac{\partial (P(\epsilon_1) P(\epsilon_2))}{\partial E}\right)_{y=y_0} \frac{d y_0}{dE}
\nonumber\\
&& {}+ (P(\epsilon_1) P(\epsilon_2))_{y=y_0} \frac{d^2 y_0}{dE^2}
\end{eqnarray}
and so on. Note that $\frac{\partial P(\epsilon)}{\partial E}= \frac{dP(\epsilon)}{d\epsilon}\frac{\partial\epsilon}{\partial E}$.

The integration term on the right is always same for both limits. The other terms are of the form $\frac{d^{m_1}P(\epsilon_1)}{d\epsilon_1^{m_1}} \frac{d^{m_2}P(\epsilon_2)}{d\epsilon_2^{m_2}}\frac{d^k y_0}{dE^k}$. Since $\frac{d^k y_0}{dE^k}$ is always discontinuous at $E=E_2$ for any $k$, a non-zero or infinite value of $\frac{d^{m_1}P(\epsilon_1)}{d\epsilon_1^{m_1}} \frac{d^{m_2}P(\epsilon_2)}{d\epsilon_2^{m_2}}$ is required for a discontinuity in $\rho$. For example, if $P(\epsilon)$ has the form $(1-2x/w)^{\lambda}$ at the edge, then $2\lambda-m_1-m_2>0$ for analyticity. Therefore, the derivative on the left side of the equation will be discontinuous at $E=E_2$ if it is of order $(n+1)$ or greater where $n$ is the smallest integer greater than or equal to $2\lambda$.

For a non-bounded distribution, $y_0$ will be infinity, independent of $E$ and therefore the DOS is always smooth.
In the 2-site problem, since the IPR and its derivatives are only a function of y and independent of x, they will have discontinuities at the same orders and energies as the DOS.

\end{document}